\documentstyle[11pt,mymorion,psfig]{article}

\newcommand{\NP}[1]{ Nucl.\ Phys.\ {\bf #1}}

\newcommand{\PL}[1]{ Phys.\ Lett.\ {\bf #1}}

\newcommand{\PR}[1]{Phys.\ Rev.\ {\bf #1}}
\newcommand{\PRL}[1]{ Phys.\ Rev.\ Lett.\ {\bf #1}}

\def\be{\begin{equation}}
\def\ee{\end{equation}}
\def\bea{\begin{eqnarray}}
\def\eea{\end{eqnarray}}

\begin{document}

\hfill{SLAC-PUB-8147}

\hfill{May 1999}

\title{NEW TESTS OF PERTURBATIVE QCD INSPIRED BY 
HYPOTHETICAL TAU LEPTONS
\footnote{Research partially supported by the Spanish 
CICYT under contract AEN97-1693
and the U.S. Department of Energy DE-AC03-76SF00515.}}
\author{J. R. PELAEZ$^1$, S.J. BRODSKY$^2$ and N. TOUMBAS$^2$}

\address{
$^1$Departamento de F\'{i}sica Te\'orica.
Universidad Complutense de Madrid.
28040 Madrid. SPAIN\\
$^2$Stanford Linear Accelerator Center. Stanford University,
Stanford, California 94309. U.S.A.
}
\maketitle
\abstracts{
Inspired by the relation between the hadronic decay of the
$\tau$ lepton and the  $e^+ e^-$
annihilation into hadrons, we derive new tests of perturbative QCD.
We design a set of
commensurate scale relations to test the
self-consistency of leading-twist QCD predictions
for any observable which defines an effective charge.
This method provides renormalization scheme and scale
invariant probes of QCD which can be applied
over wide data ranges.}
\begin{center}
Talk presented at the XXXIVth Rencontres de Moriond:\\
{\bf QCD and High Energy Hadronic Interactions}\\
Les Arcs, Bourg St. Maurice, France, March 20-27th, 1999
\end{center}

\section{Introduction}

The $\tau$ lepton hadronic width,
$R_{\tau} = \Gamma(\tau^-\rightarrow{\nu_{\tau}
+{\rm hadrons}})/\Gamma(\tau^-\rightarrow{\nu_{\tau}e^-\bar{\nu_e}})$
plays an important role in
the determination of the QCD coupling \cite{Braaten}. Its
analysis has been performed using integral moments which
minimize the sensitivity to the low energy data
\cite{Davier}. In particular, just by integrating the measured
spectral functions up $M$ we can simulate
the physics of hypothetical $\tau$ leptons\cite{Davier} with masses
$M$ smaller than the physical one.  Their hadronic widths
yield a crucial test of perturbative QCD
(PQCD), since they are related to the $e^+e^-$ annihilation
cross section into hadrons $R_{e^+e^-}$ through
\begin{equation}
R_{\tau}(M) =
\frac{2}{(\sum_f{q_f^2})}
\times\int^{M^2}_0\frac{\,d\,s}{M^2}
\left(1 -
\frac{s}{M^2}\right)^2\left(1 +
\frac{2s}{M^2}\right)R_{e^+e^-}(\sqrt{s}).
\label{defRt}
\end{equation}

In this paper we report on a recent proposal \cite{nosotros} of
self-consistency tests of PQCD, motivated by the above relations,
which can be applied to any observable which
defines an effective charge. These tests are examples of
relations between observables at two different scales, which are
called ``commensurate scale relations" \cite{CSR}.

Effective charges  are defined as the entire radiative
contribution to an observable \cite{effcha}. For instance,
assuming $f$ massless flavors, we have
\begin{equation}
R_{e^+e^-}(\sqrt{s})\equiv(3\sum_f
q_f^2)\left[1+\frac{\alpha_R(\sqrt{s})}{\pi}\right]\quad ,\quad
R_{\tau}(M)\equiv R^0_{\tau}(M)
\left[1+\frac{\alpha_\tau(M)}{\pi}\right],
\label{ReePQCD}
\end{equation}
where the effective charges $\alpha_R$ and $\alpha_\tau$ can
be written as a series
in $\alpha_s/\pi$ in any given renormalization scheme.
Their relevance is given by the fact that they
satisfy the renormalization group equation with the same
coefficients $\beta_0$ and $\beta_1$ as the usual coupling
$\alpha_s$.

At this point we can make use of the Mean Value Theorem in
eq.(\ref{defRt}), to relate $\alpha_R$ and $\alpha_\tau$ by a scale shift
\begin{equation}
\alpha_\tau(M) = \alpha_R(\sqrt{s^*}), \quad\mbox{with}\quad
\lambda_\tau=
\frac{\sqrt{s^*}}{M} = \exp\left[ -{19\over 24} - {169\over 128}
{\alpha_R(M)\over \pi} + \cdots\right],
\label{CSStau}
\end{equation}
where the value of $\lambda_\tau$ is a {\em prediction of
 NLO leading twist QCD}.
This result was first obtained in \cite{CSR} by using NNLO,
however, we will see how it is due to the fact that both effective
charges evolve with universal
$\beta_0$ and $\beta_1$ coefficients.\cite{nosotros}

\section{Tests of PQCD for a general observable}

These relations can be generalized to arbitrary observables
$O(s)$, with an associated effective
charge $\alpha_O$, by defining new effective charges
\begin{equation}
\alpha_f(M) \equiv \frac{\int^{M^2}_0 \frac{d\,s}{M^2}\,f\left(
\frac{s}{M^2}\right)\alpha_O(\sqrt{s})}{\int^{M^2}_0
\frac{d\,s}{M^2}f\left(
\frac{s}{M^2}\right)},
\label{alphaRf}
\end{equation}
where we can choose $f(x)$ to be any smooth, integrable function of
$x=s/M^2$. Once more
\begin{equation}
\alpha_f(M)=\alpha_O(\sqrt{s^*_f}),\hspace{2cm}    0\leq s^*_f\leq M^2.
\label{MVT}
\end{equation}
{\em Note that this relation only involves data for the observable
$O(s)$} and thus provides a self-consistency test for the
applicability of leading twist QCD. To obtain the relation between
the commensurate scales, we consider the running of $\alpha_O$ up to third
order
\begin{eqnarray}
\frac{\alpha_O(\sqrt{s})}{\pi}= \frac{\alpha_O(M)}{\pi} -
\frac{\beta_0}{4} \ln{\left(\frac{s}{M^2}\right)}
\left(\frac{\alpha_O(M)}{\pi}\right)^2  + \frac{1}{16}\left[
\beta_0^2 \ln^2\left(\frac{s}{M^2}\right)-
\beta_1 \ln\left(\frac{s}{M^2}\right)\right]
\left(\frac{\alpha_O(M)}{\pi}\right)^3
\ldots
\label{aR2order}
\end{eqnarray}
We substitute for $\alpha_O$
in eq. (\ref{alphaRf}) to find
\begin{eqnarray}
\frac{\alpha_f(M)}{\pi} = \frac{\alpha_O(M)}{\pi} -
\frac{\beta_0}{4}\left(\frac{I_1}{I_0}\right)
\left(\frac{\alpha_0(M)}{\pi}\right)^2
+ \frac{1}{16}\left[\beta_0^2 \left(\frac{I_2}{I_0}\right)-
\beta_1 \left(\frac{I_1}{I_0}\right)\right]
\left(\frac{\alpha_O(M)}{\pi}\right)^3
\ldots,
\label{aRf2order}
\end{eqnarray}
where  $I_l = \int^{1}_0f(x)(\ln{x})^ld\,x$. Hence
\begin{equation}
\lambda_f \equiv \frac{\sqrt{s^*_f}}{M}=\exp{\left[\frac{I_1}{2I_0} +
\frac{\beta_0}{8}\left(\left(\frac{I_1}{I_0}\right)^2 - \frac{I_2}{I_0}
\right)\frac{\alpha_O(M)}{\pi}\right]}.
\label{sstar}
\end{equation}
Note that $\lambda_f$ is
constant to leading order, and therefore $\alpha_f$ satisfies the same renormalization group
equation as $\alpha_O$ with the same coefficients $\beta_0$ and
$\beta_1$; i.e., $\alpha_f$ is an effective charge.

Note that eq.(\ref{MVT}) relates an observable with an integral over
itself. It is also possible to obtain differential
relations \cite{Carlos}, but here we will simply illustrate the use of the
integral relations.

\section{Example: self-consistency test of  $R_{e^+e^-}$ data.}

Let us then set $O=R_{e^+e^-}$. In order to suppress
the low energy region,
where non-perturbative effects are important,
we shall set $f(x)=x^k$,
with $k$ some positive number. Thus
\begin{equation}
\alpha_k(M) = \alpha_R(\lambda_k\,M)\quad \hbox{with}\quad
\lambda_k=e^{\frac{-1}{2(1+k)}},
\label{atk}
\end{equation}
When comparing with $R_{e^+e^-}$ data,
we take into account the mass effects using \cite{PQW}:
\begin{eqnarray}
R_{e^+e^-}(\sqrt{s})&=&3\sum_1^f q_i^2 \frac{v_i(3-v_i^2)}{2}\left[ 1+g(v_i)
\frac{\alpha_R(\sqrt{s})}{\pi}\right]
\equiv R_0(\sqrt{s})+R_{Sch}(\sqrt{s})\frac{\alpha_R(\sqrt{s})}{\pi}
\label{Rsch} \\
g(v)&=&\frac{4\pi}{3}\left[\frac{\pi}{2v}-\frac{3+v}{4}\left(
\frac{\pi}{2}-\frac{3}{4\pi}\right)\right]\label{g(v)}
\end{eqnarray}
where $v_i$ is the velocity of the initial
quarks in their CM frame. The $v_i(3-v_i^2)/2$ factor is the parton model
mass dependence and $g(v)$ is a QCD modification of
the Schwinger correction.
The quark masses have been taken as effective parameters which
provide a good fit to the smeared data.
All these corrections spoil eq.({\ref{atk}}), but
they are only important near the quark thresholds. At higher energies
they tend to unity and quark masses become irrelevant, which is why our
study is restricted to this regime.
\begin{figure}
\hbox{\psfig{file=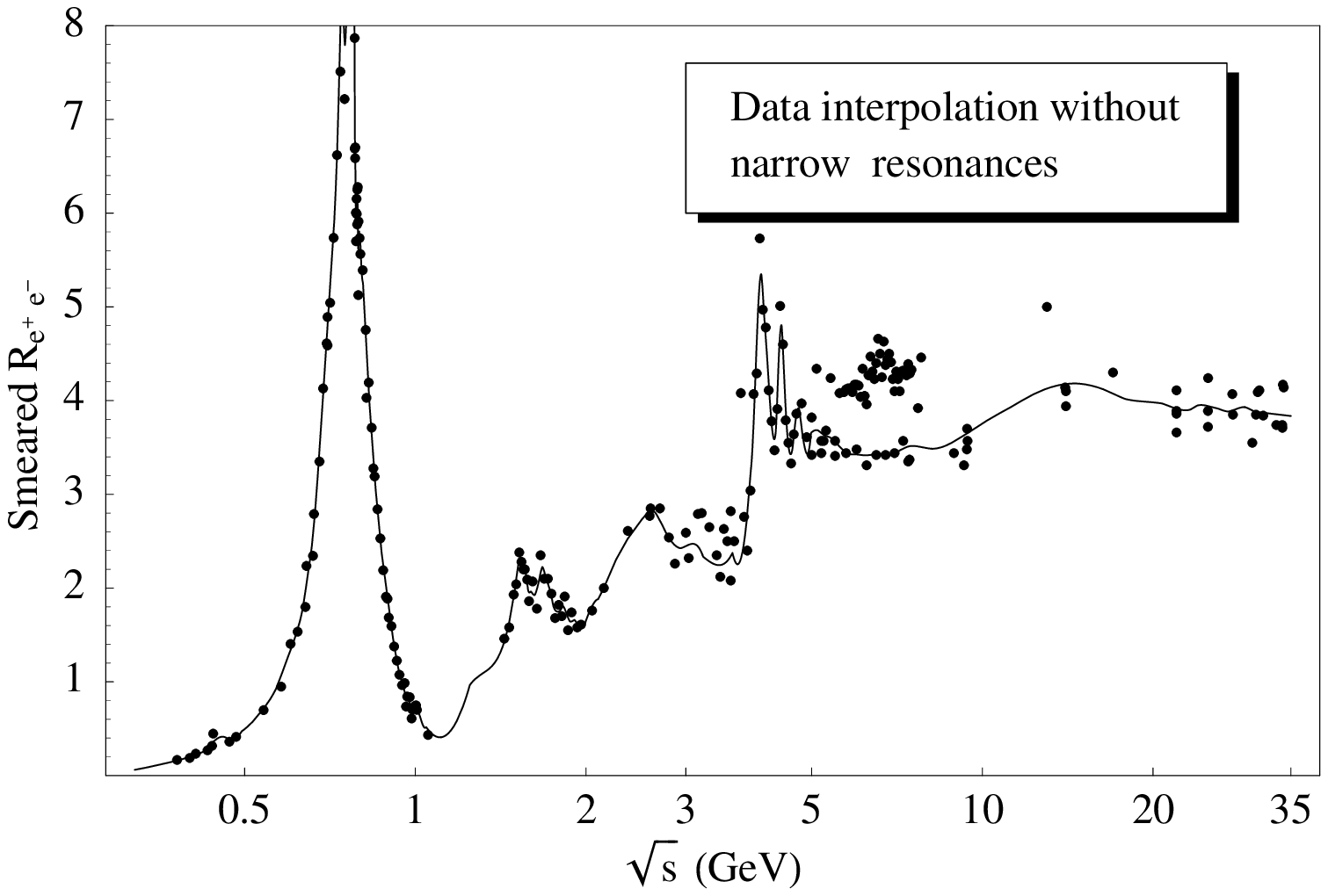,width=7.8cm}\hspace{.2cm}
\psfig{file=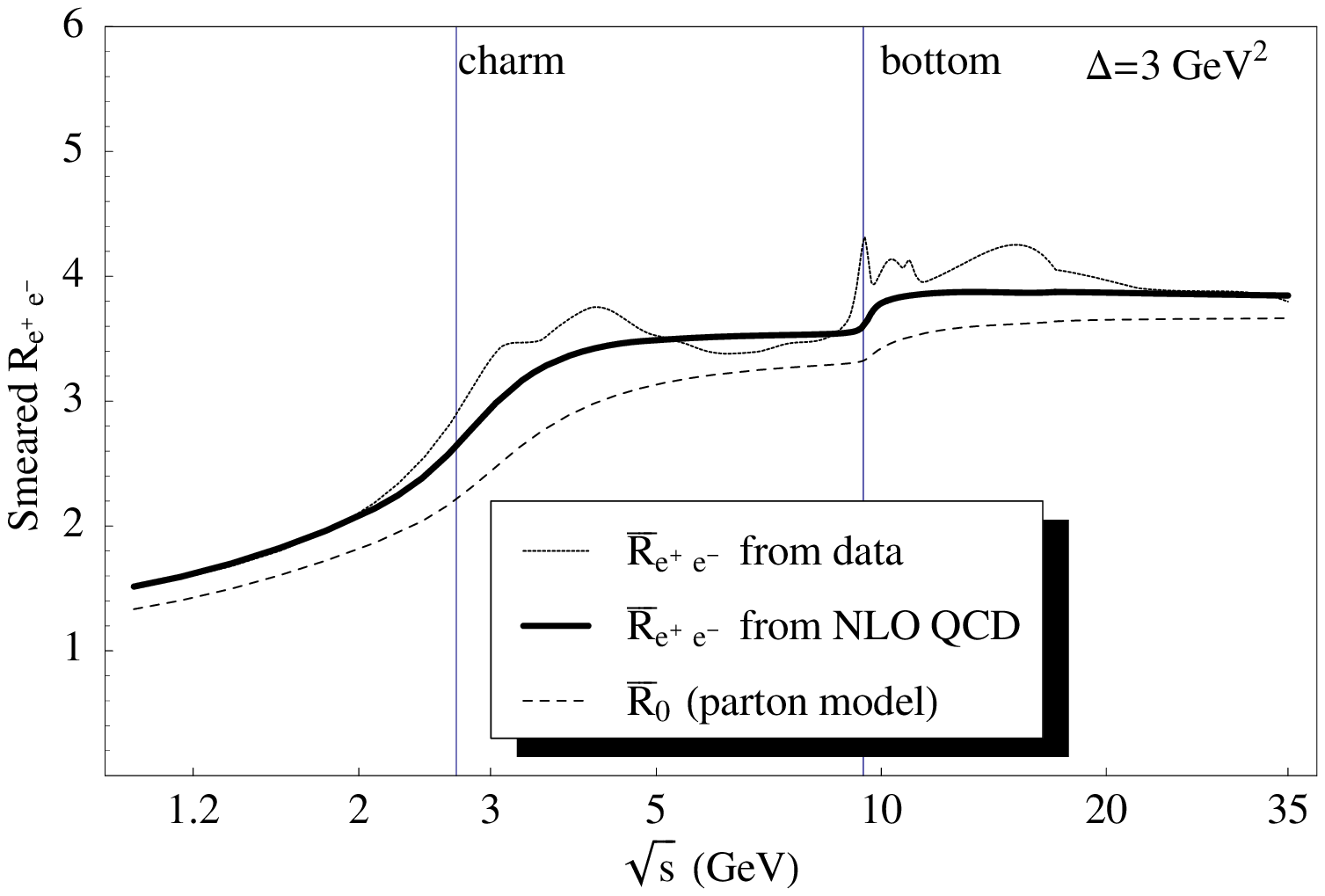,width=7.8cm}}
{\footnotesize {\bf Figure 1.a)}
Interpolation of the central values of $R_{e^+e^-}$ data (see \cite{nosotros}
for references). Note the discrepancy in the central values of experiments
 between 5 and 10 GeV.
{\bf b)} Smeared $R_{e^+e^-}$.}
\end{figure}
Still we cannot compare directly with
the data since we observe hadrons, not quarks.
Following \cite{PQW} we define smeared quantities as follows:
\begin{equation}
\bar{R}(\sqrt{s})=
\frac{\Delta}{\pi}\int^\infty_0 \frac{R(\sqrt{s'})}
{(s-s')^2-\Delta^2}\,d\,s'
\label{smear}
\end{equation}
By smearing $R_{e^+e^-}$ over a range of energy,
$\Delta E$, we focus the physics to the time $\Delta t = 1/\Delta
E $ where an analysis in terms of quarks and gluons
is appropriate.
In what follows we use the standard value $\Delta=3\,\hbox{GeV}^2$
\cite{PQW,MaSt}. The smearing effect can be seen comparing Fig.1.a,
which shows an interpolation of the $R_{e^+e^-}$ data,
(see \cite{nosotros} for references) with Fig. 1.b.
Note that {\em any fit using the QCD functional dependence will  always satisfy}
eq.(\ref{atk}) identically. To avoid this bias, we have parameterized
the narrow resonances using their Breit-Wigner form, and we have interpolated
the remaining data (see \cite{nosotros} for details).

Finally, using eqs.(\ref{Rsch}) and (\ref{smear}), we define
smeared charges:
\begin{equation}
\bar{\alpha}_R(\sqrt{s})=
\frac{\bar{R}_{e^+e^-}(\sqrt{s})-\bar{R}_0(\sqrt{s})}{\bar{R}_{Sch}(\sqrt{s})},
\end{equation}
and similarly for $\bar{\alpha}_k$.
According to the previous discussion we expect the smeared charges to satisfy
eq.(\ref{sstar}) in energy regions where the threshold corrections
can be neglected.

Thus, in Fig.2.a we compare $\bar{\alpha}_R(\sqrt{s^*})$ with
 $\bar{\alpha}_k (\sqrt{s^*}/\lambda_k)$.
The agreement for $\alpha_0$ is poor since the low energy region
is not suppressed enough. However
we find a reasonable agreement for $\alpha_1$ in several regions,
agreement which disappears if we do not shift the scales.
There are two regions of particular interest where we find a disagreement:
First, from 5 to 10 GeV where there is a well known
incompatibility between experiments (see Fig.1.a and
ref.\cite{discrepancy}). In Fig.1.a. we have kept the most recent data,
as it is standard in the literature, but still their central values
are systematically lower than the QCD predictions.
which is why eq.(\ref{atk}) does not seem to hold.
Our test correctly shows this incompatibility.
\begin{figure}
\hbox{
\psfig{file=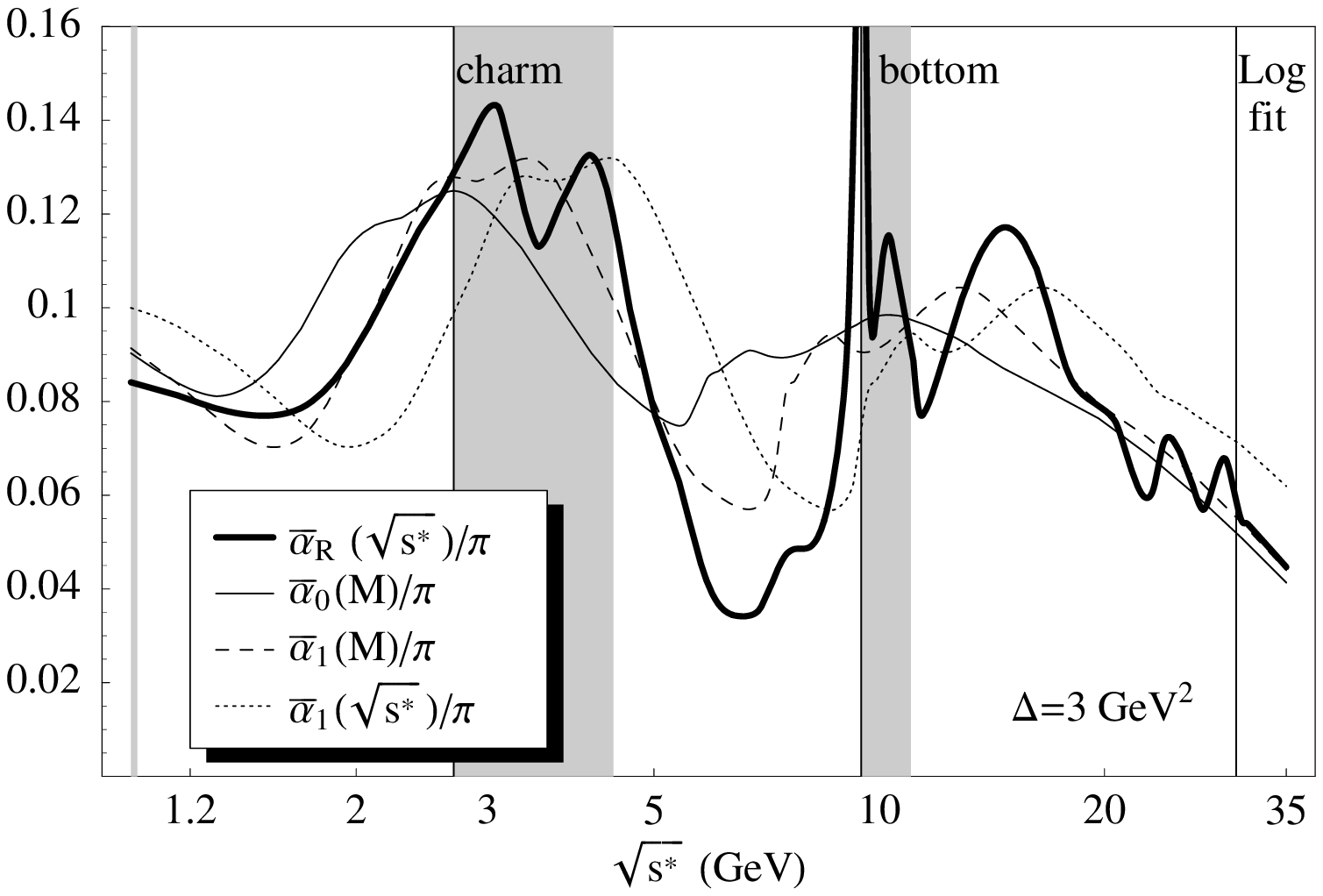,width=7.8cm}\hspace{.2cm}
\psfig{file=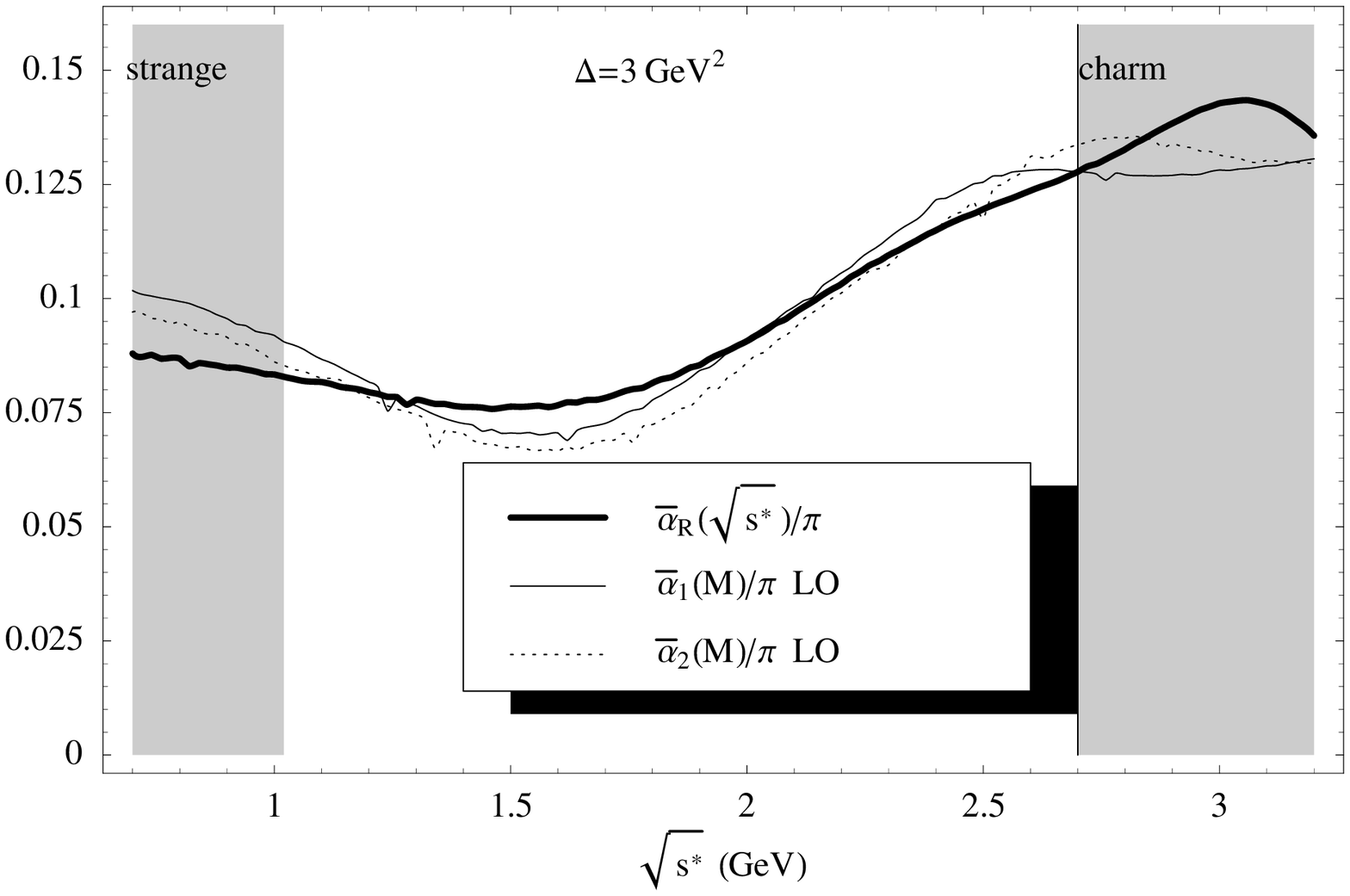,width=7.8cm}
}
{\footnotesize {\bf Figure 2.a)}
Comparison between $\bar\alpha_R(\sqrt{s^*})$ and different
$\bar{\alpha}_k$ moments at $M=\sqrt{s^*}/\lambda_k$.
The dotted line shows how the agreement is spoilt
if we do not shift $\sqrt{s^*}$ to $M$.
{\bf b)}
Comparison between $\bar\alpha_R(\sqrt{s^*})$ and different
$\bar{\alpha}_k$ moments at $M=\sqrt{s^*}/\lambda_k$
in the low energy region.
}
\end{figure}

Second, we show in Fig. 2.b., the physical
 $\tau$ region where, considering
that we are using LO QCD and central data values,
the agreement looks quite satisfactory.
This is encouraging for the applicability of PQCD
in the region near the real $\tau$ lepton.
However, at energies $\sqrt{s}\sim 1.5$ GeV,
our results support the claims that the $R_{e^+e^-}$
data could be 6-7\%
lower than expected from $R_\tau$ data\cite{Groote}.
Note, however, that our conclusions have been obtained
using {\em only data on $R_{e^+e^-}$}.

\section{Conclusions}

Motivated by the relation between $R_{e^+e^-}$ and the $R_\tau$, as well as
the ideas of commensurate scale relations, we
have presented new tests of PQCD. They can be applied in a wide
energy range to any observable which defines an effective charge, and
they are renormalization scheme and scale independent.

As an example, we have tested the self-consistency of
existing $R_{e^+e^-}$ data according to PQCD. We have found
 a good agreement in the real $\tau$ region but  incompatibilities around the
1.5 GeV region and in the range of 5 to 10 GeV, supporting previous claims
obtained by different methods. The advantage of
our approach is that it only relates the observable with itself, which
can be very useful when applied to other experiments.
\section*{Acknowledgments}
Research partially supported by the Spanish CICYT under contract AEN97-1693
and the U.S. Department of Energy DE-AC03-76SF00515. J.R.P. thanks the
SLAC theory group for their kind hospitality.

\end{document}